\documentclass[twocolumn,aps,prd,10pt,superscriptaddress,longbibliography,floatfix,citeautoscript]{revtex4-2}

\usepackage{amsmath}
\usepackage{amssymb}
\usepackage{bm}
\usepackage{dcolumn}
\usepackage{glossaries}
\usepackage{graphicx}
\usepackage{multirow}
\usepackage{fancyvrb}
\usepackage[linesnumbered,ruled]{algorithm2e}
\usepackage{siunitx}

\usepackage[usenames,dvipsnames,svgnames]{xcolor}
\usepackage{hyperref}
\hypersetup{
    pdfnewwindow=true,      
    colorlinks=true,        
    linkcolor=Blue,         
    citecolor=Blue,         
    filecolor=Blue,         
    urlcolor=Blue           
}

\usepackage{listings}	
\newacronym{ace}{ACE}{atomic cluster expansion}
\newacronym{lg}{LiTFSI/G$_{3}$}{}
\newacronym{glyme}{Glymes}{Poly(ethylene glycol)dimethyl ether}
\newacronym{litfsi}{LiTFSI}{Li$^{+}$ bis(trifluoromethylsulfonyl)azanide}
\newacronym{li}{Li$^{+}$}{}
\newacronym{tfsi}{TFSI$^{-}$}{}
\newacronym{g3}{G$_{3}$}{triglyme}
\newacronym{dft}{DFT}{density functional theory}
\newacronym{md}{MD}{molecular dynamics}
\newacronym{cmd}{CMD}{classical molecular dynamics}
\newacronym{pimd}{PIMD}{path integral molecular dynamics}
\newacronym{mlp}{MLP}{machine-learned potential}
\newacronym{nep}{NEP}{neuroevolution potential}
\newacronym{rmse}{RMSE}{root mean square error}
\newacronym{mae}{MAE}{mean absolute error}
\newacronym{snes}{SNES}{separable natural evolution strategy}
\newacronym{vdw}{vdW}{van-der-Waals}
\newacronym{nn}{NN}{neural network}
\newacronym{vasp}{VASP}{Vienna Ab initio Simulation Package}
\newacronym{rdf}{RDF}{Radial Distribution Function}
\newacronym{pca}{PCA}{principal component analysis}
\newacronym{sdc}{SDC}{Self-Diffusion Coefficients}
\newacronym{msd}{MSD}{mean squared displacement}
\newacronym{vac}{VAC}{velocity autocorrelation function}

\DeclareSIUnit\angstrom{\text{Å}}
\DeclareSIUnit{\atom}{atom}
\DeclareSIUnit{\step}{step}
\DeclareSIUnit{\atomstepsecond}{\atom\step\per\second}

\newcolumntype{d}{D{.}{.}{-1}}

\begin{document}

\title{Structural and transport properties of LiTFSI/G$_{3}$ electrolyte with machine-learned molecular dynamics}

\author{Chenyang Cao}
\affiliation{Beijing Advanced Innovation Center for Materials Genome Engineering, Department of Physics, University of Science and Technology Beijing, Beijing 100083, China}

\author{Liyi Bai}
\affiliation{Suzhou Laboratory, Suzhou 215100, China}

\author{Shuo Cao}
\affiliation{Beijing Advanced Innovation Center for Materials Genome Engineering, Corrosion and Protection Center, University of Science and Technology Beijing, Beijing, 100083, China}

\author{Ye Su}
\affiliation{Beijing Advanced Innovation Center for Materials Genome Engineering, Department of Physics, University of Science and Technology Beijing, Beijing 100083, China}

\author{Yanzhou Wang}
\affiliation{Department of Applied Physics, QTF Center of Excellence, Aalto University, FIN-00076 Aalto, Espoo, Finland}

\author{Zheyong Fan}
\affiliation{College of Physical Science and Technology, Bohai University, Jinzhou 121013, P. R. China}

\author{Ping Qian}
\email{qianping@ustb.edu.cn}
\affiliation{Beijing Advanced Innovation Center for Materials Genome Engineering, Department of Physics, University of Science and Technology Beijing, Beijing 100083, China}

\date{\today}
\begin{abstract}
The lithium bis(trifluoromethylsulfonyl)azanide-triglyme (LiTFSI/G3) electrolyte plays a critical role in the performance of lithium-ion batteries. However, its solvation structure and transport properties at the atomic scale remain incompletely understood. In this study, we develop an efficient and accurate neuroevolution potential (NEP) model by integrating bootstrap and active learning strategies. Using machine-learned NEP-driven molecular dynamics simulations, we explore the structural and diffusion properties of LiTFSI/G3 across a wide range of the solute-to-solvent ratios, systematically analyzing electrolyte density, ion coordination, viscosity, and lithium self-diffusion. The computed densities show excellent agreement with experimental data, and pair correlation analysis reveals significant interactions between lithium ions and surrounding oxygen atoms, which strongly impacts Li$^+$ mobility. Viscosity and diffusion calculations further demonstrate that increasing LiTFSI concentration enhances Li-O interactions, resulting in higher viscosity and reduced lithium diffusion. Additionally, machine learning-based path integral molecular dynamics (PIMD) simulations confirm the negligible impact of quantum effects on Li$^+$ transport. The electrolyte-specific protocol developed in this work provides a systematic framework for constructing high-fidelity machine learning potentials for complex systems.
\end{abstract}
\maketitle

\section{Introduction}

Lithium-ion batteries (LIBs) are essential for modern energy storage, particularly in electric vehicles and portable electronics, due to their high energy density, long cycle life~\cite{2024MD_2}, and reliability~\cite{wulandari2023lithium,2017LiTFSI_1}. The electrolyte, a key component, governs ion transport and electrochemical stability~\cite{2024G3_1}, directly influencing battery performance and lifespan \cite{LU2024103741, gauthier2015electrode,2021LiTFSI_3}. Conventional liquid electrolytes, typically consisting of organic solvents (e.g., carbonates) and lithium salts (e.g., \gls{litfsi})~\cite{2017LiTFSI_2}, offer high ionic conductivity and electrode compatibility. However, their volatility, flammability, and limited electrochemical stability present significant challenges, necessitating the development of safer and more efficient alternatives.

\Gls{md} simulations provide atomic-scale insights into electrolyte structure~\cite{C8CP06214E,2024MD_1,B612297C,2024MD_3} and transport~\cite{2024MD_2}, revealing key mechanisms of ion solvation, diffusion, and viscosity. However, traditional empirical force fields often fail to accurately capture complex intermolecular interactions, leading to discrepancies in predicting key transport properties. \Gls{mlp}, trained on first-principles data, has emerged as a powerful alternative, achieving near-\gls{dft} accuracy while maintaining computational efficiency. Successfully applied in metals, catalysis~\cite{Yang23p829}, and electrolytes~\cite{D0CC03512B,doi:10.1021/acs.jctc.2c00926,wang2025chem,2024MLP_1}, \gls{mlp} leverages large datasets and neural networks to accurately model atomic interactions, providing a robust framework for high-fidelity electrolyte simulations.

In this study, we develop a high-accuracy \gls{nep} \cite{PhysRevB.104.104309, fan2022jpcm, 10.1063/5.0106617, song2024general} for the \gls{litfsi}--\gls{g3} electrolyte using an active learning approach. \Gls{nep} is a highly efficient machine-learned potential approach with widespread applications in modeling complex materials \cite{Ying2025cpr}, including ionic diffusion in solid-state electrolytes \cite{Yan2024CM}. The \gls{lg} system was selected for its significant influence on \gls{li} transport and its favorable electrochemical properties, including high solubility and stability \cite{ZHANG20061379, doi:10.1021/ja203983r}. Specifically, \gls{g3}, as a solvent, exhibits low volatility, non-flammability, and strong solvation capability, making it widely used in metal-ion batteries \cite{C9TA03261D, 10.1063/5.0214769}. Meanwhile, \gls{litfsi}, a common lithium salt, strongly dissociates in non-aqueous solvents, with its \gls{tfsi} anion reducing viscosity and enhancing ionic conductivity \cite{LEWANDOWSKI2009601}.

Using this \gls{nep} potential, we systematically investigate the structural and transport properties of \gls{lg} electrolytes, including density, ion coordination, viscosity, and \gls{li} self-diffusion. Our results demonstrate that \gls{cmd} accurately reproduces experimental density trends, and \gls{pimd} \cite{ying2025jcp} confirms the negligible impact of quantum effects on \gls{li} transport. Additionally, increasing \gls{litfsi} concentration strengthens ion-ion interactions, leading to higher viscosity and reduced \gls{li} diffusion.

\section{Models and Methods}

\subsection{Datasets for training the NEP model}

\begin{figure*}
\centering
\includegraphics[width=1.8\columnwidth]{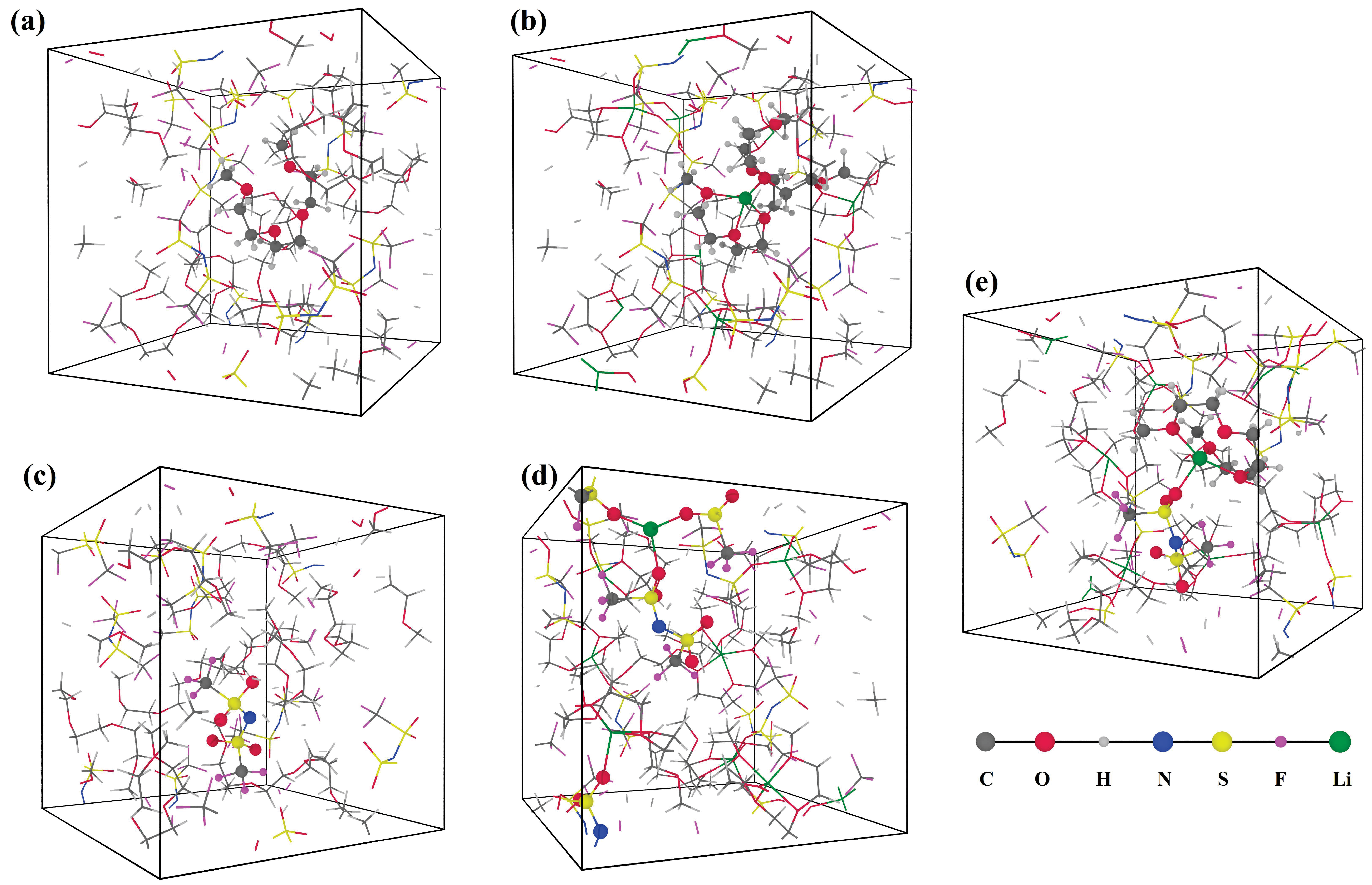}
\caption{Snapshots of a 460-atom electrolyte solution system with a solute-to-solvent ratio of \gls{lg}~$=1$.  
The ball-and-stick model highlights: (a) one \gls{g3} molecule, (b) one \gls{li} cation bonded to four oxygen atoms from two neighboring \gls{g3} molecules, (c) one \gls{tfsi} anion, (d) one \gls{li} cation bonded to three oxygen atoms from two neighboring \gls{tfsi} anions, and (e) one \gls{li} cation bonded to one oxygen atoms from one \gls{tfsi} anion and to two oxygen atoms from one solvent \gls{g3} molecule. The Li-O bonding is cut within 2.3 ~\AA. C, O, H, N, S, F, and Li atoms are represented by dark grey, red, light grey, blue, yellow, pink, and green spheres, respectively.}
\label{fig:model}
\end{figure*}

We utilize a publicly available, well-established dataset for \gls{lg}~\cite{WANG2023100061}, constructed from extensive MD simulations of the electrolyte solution. This dataset spans six solute-to-solvent molar fractions: \gls{lg}$~= 0$, $0.2$, $0.36$, $0.58$, $0.82$, and $1$. Configurations were sampled from \gls{cmd} trajectories performed in the NPT ensemble at ambient conditions and in the NVT ensemble over a temperature range of $300$–$500$ K for each molar fraction, yielding a total of $4500$ \gls{lg} configurations.  Figure~\ref{fig:model} illustrates a representative snapshot of a $460$-atom electrolyte solution system with a solute-to-solvent ratio of \gls{lg}~$=1$. The ball-and-stick model highlights the key molecular components: solvent \gls{g3} and solute \gls{litfsi} molecules. Due to the high degree of atomic environment overlap resulting from dense MD sampling, reducing the training set size is essential to enhance data efficiency while preserving configurational diversity.  

\begin{figure}
\centering
\includegraphics[width=1\columnwidth]{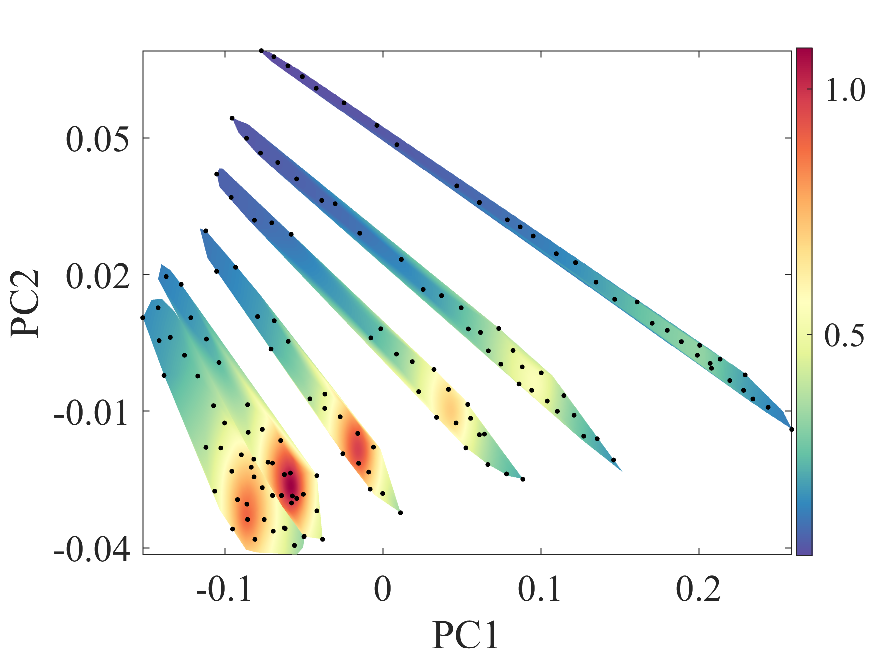}
\caption{Principal component analysis (PCA) of the public electrolyte dataset, comprising six \gls{lg} molar fractions, as reported by Wang \textit{et al.}~\cite{WANG2023100061}. The analysis was performed using the \textsc{PyNEP} code~\cite{PyNEP}. The two principal components are visualized with a color bar, where higher values indicate greater projection density. Black dots denote $154$ sparsely selected structures, ensuring configurational diversity and data efficiency.}
\label{fig:pca}
\end{figure}

To address this, we applied principal component analysis (PCA) to all configurations in the public dataset using the \textsc{PyNEP} code~\cite{PyNEP}, reducing the high-dimensional structural data to two principal components, as shown in Fig.~\ref{fig:pca}. The PCA projection of the original dataset, derived from six molar ratios, is visualized using a color bar. We then performed sparsification on these two components and uniformly selected $154$ sparse \gls{lg} configurations (black dots in Fig.~\ref{fig:pca}).\par

Moreover, Wang \textit{et al.} extracted \gls{lg} structures from short-time MD trajectories (e.g., 10 ps) for their dataset~\cite{WANG2023100061}, which constrained the exploration of configurational space. To overcome this limitation, we initially trained a preliminary NEP model on the $154$ configurations. Leveraging the computational efficiency of NEP, we conducted a longer 10 ns NEP-driven MD simulation, extracting hundreds of frames at equal intervals. By comparing NEP-predicted atomic forces with DFT-calculated forces, we identified and selected $106$ frames with the maximum atomic force difference exceeding $0.8$ eV/\AA~ to enhance the sparse dataset.

To further enhance the robustness of the NEP model, we expanded the mass density range by systematically varying the electrolyte box dimensions. The mass density of \gls{lg} at room temperature and pressure was adjusted triaxially from $-5\%$ to $5\%$, corresponding to a density range of $85\%$ to $115\%$, yielding $70$ new density-modulated \gls{lg} configurations. To account for quantum dynamical effects in electrolyte simulations, we performed \gls{pimd} simulations for $1.2$ ns at room temperature and pressure for each molar fraction. From these trajectories, we selected $160$ configurations by filtering frames with the maximum atomic force difference exceeding $0.8$ eV/\AA~ relative to DFT single-point calculations. In total, our training set consists of $490$ configurations encompassing $238,658$ atomic environments.  

To assess the NEP model's robustness in both interpolation and extrapolation, we constructed a test set comprising $126$ electrolyte structures. This set includes $60$ frames sampled from 10 ns NPT-CMD trajectories and $66$ frames from $1.2$ ns PIMD trajectories, all at room temperature and pressure.  

Finally, we performed single-point density functional theory (DFT) calculations on all structures in both datasets using the VASP package~\cite{Kresse1996prb}. The electronic structure was described using the projector augmented wave (PAW) method~\cite{paw1,paw2} combined with the generalized gradient approximation (GGA) parameterized by Perdew, Burke, and Ernzerhof (PBE)~\cite{Perdew1996prl}. A single Gamma $k$-point was employed for Brillouin zone sampling, with Gaussian smearing applied at a width of $0.05$ eV. The kinetic energy cutoff for plane wave expansion was set to $600$ eV, and the energy convergence threshold was fixed at $10^{-6}$ eV.

While \gls{vdw} interactions are essential for accurately describing the structural properties of the \gls{lg} electrolyte~\cite{WANG2023100061}, they were intentionally excluded from our DFT calculations. This strategy aligns with our research design, where MD simulations are conducted using the integrated NEP-D3 ~\cite{Ying_2024_jpcm} approach by combining NEP and the D3  dispersion correction \cite{10.1063/1.3382344} to explicitly handle \gls{vdw} effects. By implementing long-range dispersion corrections directly within the MD simulations, we can achieve precise modeling of \gls{vdw} interactions without complicating the construction of the NEP model.

\subsection{The NEP Formalism}  

The \gls{nep} framework~\cite{PhysRevB.104.104309,fan2022jpcm,10.1063/5.0106617, song2024general} combines a neural network (NN) architecture with a separable natural evolution strategy (SNES)~\cite{Schaul2011} to represent the potential energy surface and iteratively optimize model training. It employs a feedforward single-hidden-layer NN to construct a local descriptor for a central atom~$i$, with its site energy $U_i$ expressed as:  

\begin{equation}
\label{equation:u_i}
U_{i} = \sum_{\mu=1}^{N_{\rm neu}} \omega_\mu^{(1)} \tanh\left(\sum_{\nu=1}^{N_{\rm des}}\omega_{\mu\nu}^{(0)}q_{\nu}^i-b_{\mu}^{(0)}\right)-b^{(1)}.
\end{equation}
Here, $q_{\nu}^i$ represents the $\nu$-th descriptor component of atom $i$, and $\tanh(x)$ is the activation function in the hidden layer. The parameters $\omega_{\mu\nu}^{(0)}$ and $b_{\mu}^{(0)}$ define the regression matrix and bias, mapping the $\nu$-th descriptor component to the $\mu$-th neuron in the hidden layer. Similarly, $\omega_\mu^{(1)}$ and $b^{(1)}$ map the hidden layer output to $U_i$. The numbers of input layer descriptors and hidden layer neurons are denoted as $N_\mathrm{des}$ and $N_\mathrm{neu}$, respectively.

The descriptor vector $\mathbf{q}^i$ in \gls{nep} typically comprises radial (2-body) and angular (many-body) components, conceptually similar to Behler-Parrinello symmetry functions~\cite{behler2007prl} and the optimized smooth overlap of atomic positions (SOAP)~\cite{bartok_2013,2019_Miguel_soap}. The 2-body terms $q_n^i$ are expressed as a sum of radial basis functions $g_n(r_{ij})$:

\begin{equation}
\label{equation:rad_des}
q_{n}^i = \sum_{j\neq{i}}g_n(r_{ij}),~0 \leq {n} \leq n_{\rm max}^{\rm R}.
\end{equation}
Here, $r_{ij}$ is the distance between atoms $i$ and $j$, and each $g_n(r_{ij})$ is expanded as a linear combination of Chebyshev polynomials.

The angular descriptor components capture 3-body and 4-body correlations using spherical harmonics. The 3-body terms $q_{nl}^i$ $(0 \leq {n} \leq n_{\rm max}^{\rm A}$,  $1 \leq {l} \leq l_{\rm max})$ are defined as:

\begin{equation}
\label{equation:ang_des}
q_{nl}^i =\frac{2l+1}{4\pi}\sum_{j\neq{i}}\sum_{k\neq{i}}g_n(r_{ij})g_n(r_{ik})P_l(\cos\theta_{ijk}),
\end{equation}
where $\theta_{ijk}$ denotes the angle between atomic pairs $ij$ and $ik$, and $P_l$ represents the Legendre polynomial. In \gls{nep} implementations, the radial functions used in 3-body components may have a distinct cutoff radius $r_{\rm c}^{\rm A}$ from that in 2-body components $r_{\rm c}^{\rm R}$.

\subsection{Trained NEP model}
\begin{figure*}
\centering
\includegraphics[width=2\columnwidth]{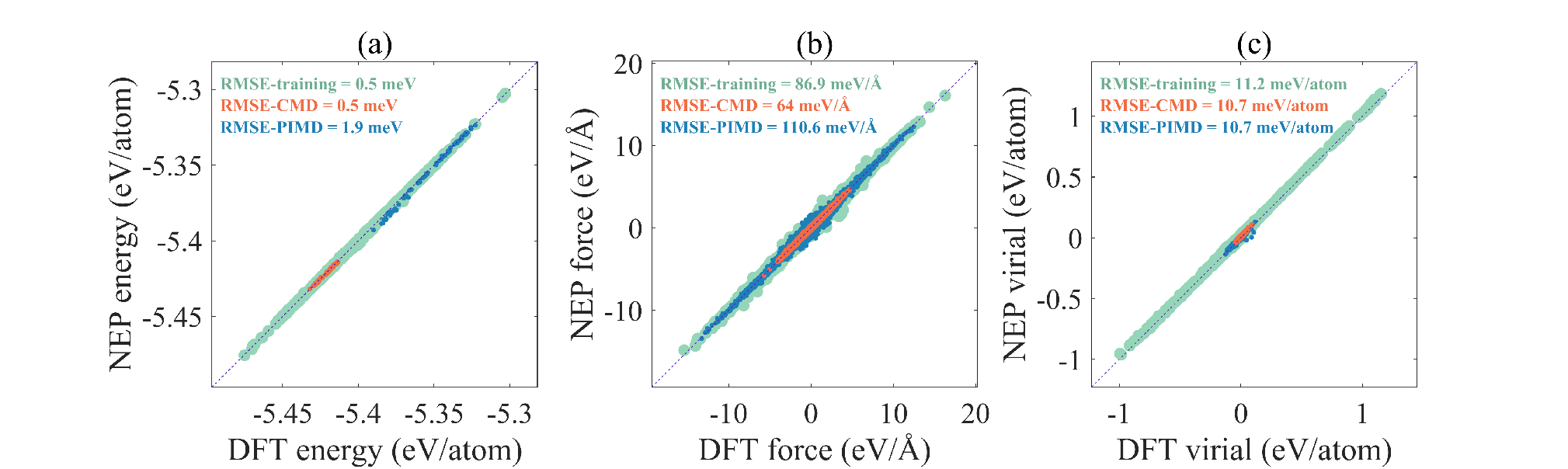}
\caption{Parities of (a) energies, (b) forces, and (c) virials predicted by \gls{nep} as against \gls{dft} reference values. The training set ($490$ configurations) is shown in light green, while the test sets include $60$ \gls{cmd} configurations (orange) and $66$ \gls{pimd} configurations (blue). The \gls{rmse} values for each dataset are provided in the corresponding panels.}

\label{fig:energy}
\end{figure*}

\begin{table}[htb]
\centering
\setlength{\tabcolsep}{2.5Mm}
\caption{Hyperparameters used for training the \gls{nep} model on \gls{lg}.}
\label{table:hyper}
\begin{tabular}{lllllllll}
\hline
\hline
Parameter & Value & Parameter & Value\\
\hline
$r_{\rm c}^{\rm R}$ & 6 \AA & $r_{\rm c}^{\rm A}$ & 6 \AA \\
$n_{\rm max}^{\rm R}$ & 10 & $n_{\rm max}^{\rm A}$ & 10 \\
$N_{\rm bas}^{\rm R}$ & 8 & $N_{\rm bas}^{\rm A}$ & 8 \\
$l_{\rm max}^{\rm 3b}$ & 4 & $l_{\rm max}^{\rm 4b}$ & 2 \\
$N_{\rm neu}$ & 30 &  $N_{\rm gen}$ & $1.3 \times 10^6$ \\
$\lambda_{\rm e}$ & 1.0  & $\mathrm{ZBL}$ & 2.5 \AA \\
$\lambda_{\rm f}$ & 1.0 & $\lambda_{\rm v}$ & 0.1 \\
$N_{\rm bat}$ & 490 & $N_{\rm pop}$ & 50 \\
\hline
\hline
\end{tabular}
\end{table}

The trainable parameters are optimized using SNES~\cite{Schaul2011} to minimize a loss function comprising a weighted sum of the root-mean-square errors (RMSEs) of energy, force, and virial stress over $N_{\rm gen}$ generations with a population size of $N_{\rm pop}$. The radial ($r_\mathrm{c}^\mathrm{R}$) and angular ($r_\mathrm{c}^\mathrm{A}$) cutoffs influence the accuracy of the NEP model. After careful testing, we selected $r_\mathrm{c}^\mathrm{R} = r_\mathrm{c}^\mathrm{A} = 6$~\AA~ as the optimal choice. To capture many-body interactions in the \gls{lg} system, we incorporated a single 4-body descriptor with $l_\mathrm{max}^\mathrm{4b} = 2$. Additionally, the universal pair potential by Ziegler, Biersack, and Littmark (ZBL)~\cite{Ziegler1985} was applied with a cutoff of $2.5$~\AA~ to enhance short-range repulsion, improving the robustness of the trained model. We employed the NEP4 version~\cite{song2024general} to train our model using a full-batch approach. All hyperparameters used in training are listed in Table~\ref{table:hyper}.

Figure~\ref{fig:energy} presents the parity plots for physical quantities, including energy, force, and virial stress, as predicted by the trained \gls{nep} model against \gls{dft} references. The results indicate that our \gls{nep} model for \gls{lg} achieves near-quantum mechanical accuracy in both the training set (light green dots) and the test set, which consists of $60$ CMD configurations (orange dots) and $66$ PIMD configurations (blue dots). Specifically, the reported \gls{rmse}s for the training set are $0.5$ meV/atom for energy (Fig.~\ref{fig:energy}(a)), $86.9$ meV/\AA~ for force (Fig.~\ref{fig:energy}(b)), and $11.2$ meV/atom for virial stress (Fig.~\ref{fig:energy}(c)).

While the \gls{cmd} test set exhibits accuracy comparable to the training set (and even surpasses it in force prediction, with an \gls{rmse} of $64$ meV/\AA~ compared to $86.9$ meV/\AA~ for the training set), the \gls{pimd} test set shows relatively lower accuracy, with a larger energy \gls{rmse} of 1.9 meV/atom and a higher force \gls{rmse} of 110.6 meV/\AA~. The higher \gls{rmse} values in the \gls{pimd} test set may be attributed to the broader configurational diversity in \gls{pimd} simulations, which exhibit a wider energy and force distribution in Fig.~\ref{fig:energy}(a-b) compared to the \gls{cmd} test set.

\subsection{Computational details}

In the \gls{lg} electrolyte system, the solute concentration of \gls{litfsi} plays a crucial role in the diffusion behavior of \gls{li}. To investigate this effect, we systematically studied electrolyte systems across a broad range of solute molar ratios, including \gls{lg}~$=0$ (pure \gls{g3}), $0.0625$, $0.20$, $0.36$, $0.58$, $0.82$, and $1$. Each system consists of approximately $500$ atoms. A structural snapshot of the solute-to-solvent molar ratio \gls{litfsi}/\gls{g3}~$=1$ is shown in Fig.~\ref{fig:model}. All molecular dynamics (MD) simulations in this study were performed using $3\times3\times3$ supercells, with atom counts ranging from $12,000$ to $15,000$. Detailed structural information for \gls{litfsi}/\gls{g3} supercells at different molar ratios is provided in Table~\ref{table:MD}.

\begin{table}[htb]
\setlength{\tabcolsep}{1.0Mm}
\caption{Structural details of electrolyte supercells with different \gls{lg} molar ratios at $300$~K, including system size (nm), number of atoms ($N_\mathrm{atoms}$), number of solute \gls{litfsi} molecules ($N_{\mathrm{\gls{litfsi}}}$), number of solvent \gls{g3} molecules ($N_{\mathrm{G_3}}$), \gls{li} concentration ($c$(Li$^+$) in mol L$^{-1}$), and mass densities (kg L$^{-1}$) from this work and available experimental data \cite{doi:10.1021/jp307378j}.}
\label{table:MD}
\begin{tabular}{cccccccccc}
\hline
\hline
\gls{litfsi}/\gls{g3} & Size  & $N_{\mathrm{atoms}}$ & $N_{\mathrm{G_3}}$ & $N_{\mathrm{\gls{litfsi}}}$ & $c$(Li$^+$)  & $\rho_{\mathrm{MD}}$ & $\rho_{\mathrm{Exp}}$ \\ 
\hline
0 &     52.89  & 14,580  &486 &0 &0    &0.999  &0.986\\ 
0.0625& 51.75  & 13,392  &432 &27 &0.319 &1.044 &--\\
0.20 & 52.20  & 13,446  &405 &81 &0.849 &1.137 &--\\
0.36 & 52.80 & 13,500 &378 &135 &1.279  &1.222 &--\\
0.58 & 52.61 & 12,744  &324 &189 &1.682 &1.328 &--\\
0.82 & 53.28  & 12,798 &297 &243 &1.982 &1.400 &--\\
1.00 & 53.63  & 12,420  &270 &270 &2.149  &1.414 &1.420\\
\hline
\hline
\end{tabular}
\end{table}

Using classical molecular dynamics (CMD) and path-integral molecular dynamics (PIMD), we systematically calculated various electrolyte properties at different concentrations and temperatures, including density, coordination environment, viscosity, and self-diffusion coefficients. For density calculations, electrolyte structures were equilibrated for $2$ ns in the NPT ensemble using a stochastic cell rescaling thermostat~\cite{PhysRevE.75.056707}, with density values averaged over the final $1$ ns of the trajectory. \gls{pimd} simulations included $1$ ns of equilibration followed by $0.5$ ns of statistical averaging. Pair correlation functions (PCFs) were computed from the equilibrated trajectories.  

Viscosity was determined by averaging over $30$ independent simulations, each consisting of $2$ ns of NPT equilibration followed by $1$ ns of NVE simulation with a correlation time of $200$ ps. Self-diffusion coefficients were averaged over three independent runs, each comprising $2$ ns of NPT equilibration and $2$ ns of NVT simulation using a Nosé-Hoover chain thermostat, with a correlation time of $20$ ps. For PIMD simulations, after $0.5$ ns of NPT equilibration, an additional $0.5$ ns was used for property calculations. The computation of the velocity autocorrelation function (VAC) in \gls{pimd} followed the same procedure as in \gls{cmd}.  

All calculations based on the NEP model were performed using the \textsc{GPUMD} package~\cite{fan2017cpc}, while DP model simulations were conducted in \textsc{LAMMPS}~\cite{LAMMPS2022cpc}. The time step was set to $1$ fs for \gls{cmd} and $0.5$ fs for \gls{pimd}. All MD simulations incorporated D3 dispersion corrections.

\section{Results and discussion}

\subsection{Electrolyte structure}

\gls{litfsi} serves as the electrolyte salt, and \gls{g3} as the solvent, both of which possess symmetrical chain-like structures. In the \gls{litfsi} molecular chain, the terminal CF$_3$ groups are symmetrically positioned and covalently bonded to sulfur atoms (depicted as larger orange atoms in Fig.~\ref{fig:model}(c)). Each sulfur atom is further doubly bonded to two oxygen atoms (shown as red spheres), which are located on either side of the molecular backbone. The two symmetric CF$_3$SO$_2$ groups are collectively bonded to a central nitrogen atom. In pure \gls{litfsi} molecules, Li atoms preferentially reside near the central nitrogen atom, coordinating with the surrounding oxygen atoms.

The \gls{g3} molecule adopts a long chain-like structure composed of carbon, oxygen, and hydrogen atoms, as illustrated in Fig.~\ref{fig:model}(a). The molecular chain is terminated by a methoxy (CH$_3$O-) group at one end and a methyl (CH$_3$-) group at the other, while the backbone consists of three repeating CH$_2$CH$_2$O- units.  

As shown in Fig.~\ref{fig:model}(b), \gls{li} ions (marked in green) coordinate with four oxygen atoms and are surrounded by \gls{g3} molecules. Notably, these oxygen atoms originate from both individual \gls{g3} molecules and the surrounding \gls{g3} electrolyte. Specifically, \gls{li} can coordinate with oxygen atoms from multiple \gls{g3} molecules and may also interact with oxygen atoms in \gls{tfsi} anions, as depicted in Fig.~\ref{fig:model}(d-e). This multi-coordination behavior gives rise to a more complex local chemical environment for \gls{li}, significantly influencing its solvation structure and dynamic diffusion behavior.

\subsection{Density}

\begin{figure*}
\centering
\includegraphics[width=1.8\columnwidth]{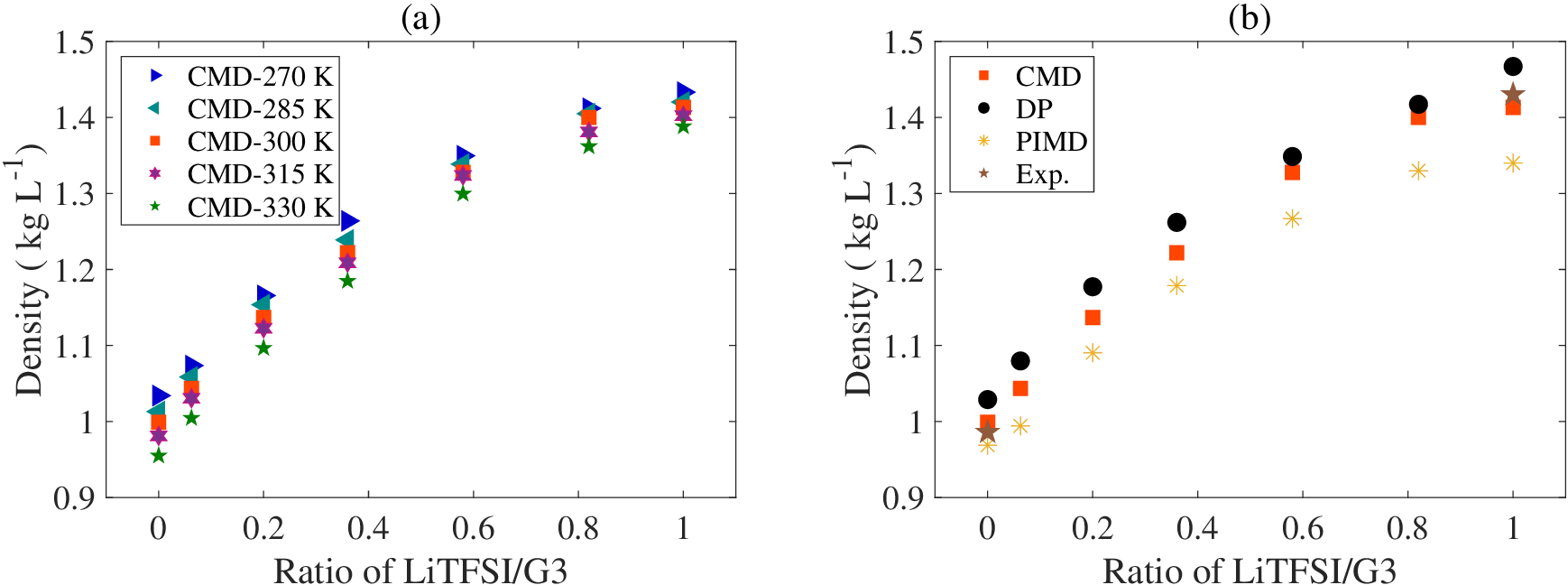}
\caption{Density as a function of \gls{lg} molar fraction ($0$–$1$): (a) CMD predictions over a temperature range of $270$–$330$ K, and (b) comparison of NEP-driven PIMD, NEP-driven CMD, and DP-driven CMD~\cite{WANG2023100061} calculations with experimental measurements~\cite{Conesa19981343,doi:10.1021/jp307378j} at $300$ K.}
\label{fig:density}
\end{figure*}

Both our tests and the study by Wang et al.~\cite{WANG2023100061} have emphasized the critical role of \gls{vdw} interactions in the strongly polarized \gls{lg} system, where long-range dispersion forces are essential for accurately predicting structural properties such as density. Under CMD calculations with D3 corrections, Fig.~\ref{fig:density}(a) presents the density as a function of the mole fraction of \gls{lg} at room temperature. Since the electrolyte \gls{litfsi} is denser than the solvent \gls{g3}, an increase in \gls{lg} concentration intuitively leads to higher density. Additionally, due to positive thermal expansion, higher temperatures result in lower-density structures.

Due to quantum effects in the electrolyte, \gls{pimd} simulations were performed. \gls{pimd} employs a quantum mechanical formulation in which quantum particles are represented as "strings" or "paths" composed of classical particles. By leveraging the path integral approach, \gls{pimd} effectively captures quantum behavior, providing a more accurate representation of quantum effects~\cite{10.1063/1.3489925,10.1063/5.0050450,10.1063/1.5005059}. This method is particularly effective in capturing quantum fluctuations, making it a valuable tool for studying quantum systems. However, compared to \gls{cmd}, \gls{pimd} entails significantly higher computational costs due to the additional degrees of freedom introduced by path integration. To balance computational accuracy and efficiency, we employ 32 "strings" to represent each particle in this study.

Figure~\ref{fig:density}(b) presents the density-concentration relationship for \gls{lg} solutions at room temperature, as calculated using \gls{cmd} and \gls{pimd}. For comparison, data from DP-driven CMD simulations~\cite{WANG2023100061} and experimental measurements~\cite{Conesa19981343,doi:10.1021/jp307378j} are also included. The results show that our NEP-driven CMD calculations are in good agreement with DP simulations and exhibit even better consistency with experimental data at \gls{lg} ratios of $0$ and $1$. Furthermore, due to quantum fluctuations, \gls{pimd} predicts lower densities compared to \gls{cmd} while still maintaining good agreement with experimental results.

\subsection{Pair correlation function}

\begin{figure*}
\centering
\includegraphics[width=1.8\columnwidth]{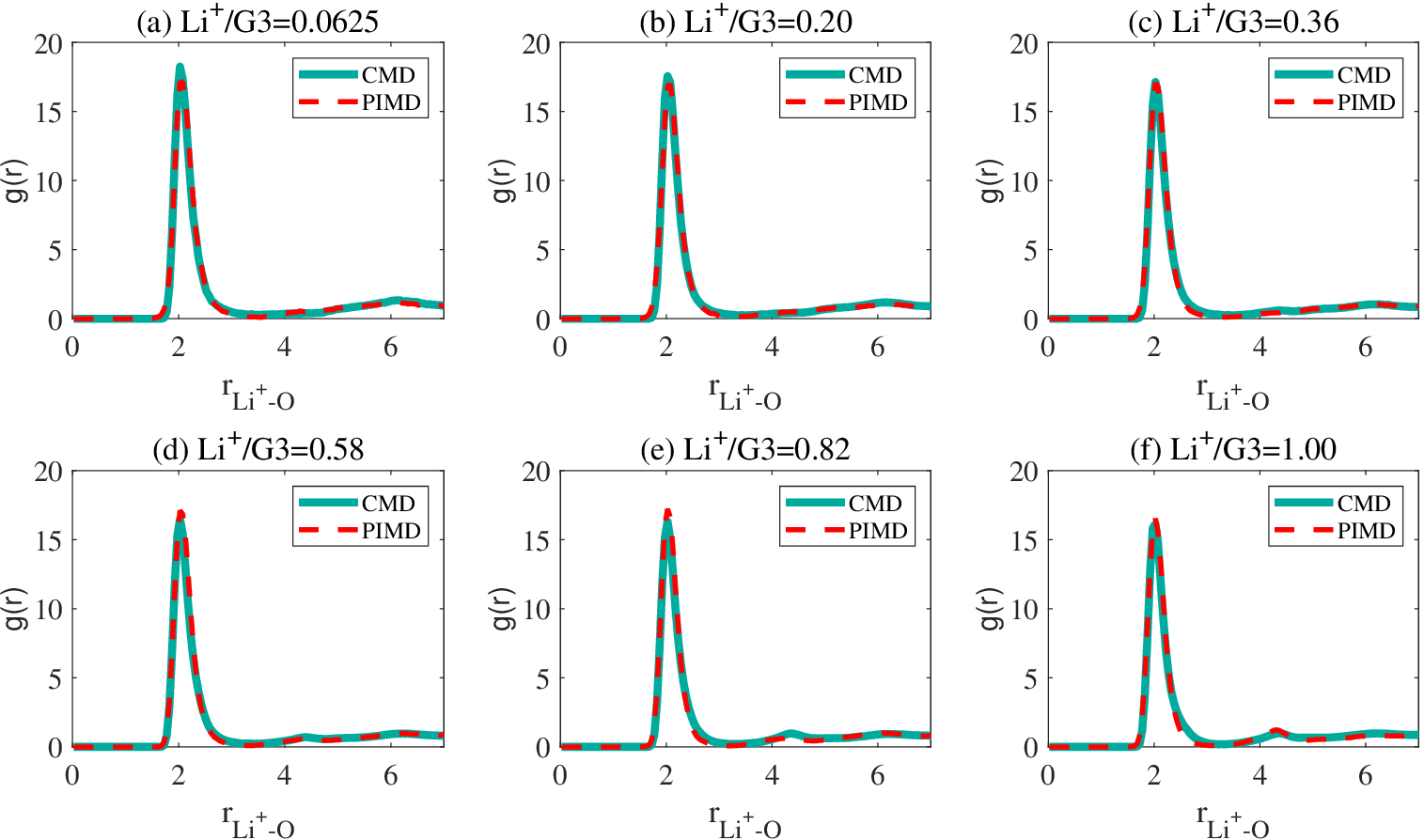}
\caption{Pair correlation function $g(r)$ for Li–O interactions in \gls{lg} at $300$ K, shown as a function of Li–O distance for different \gls{li}/\gls{g3} molar fractions: (a) $0.0625$, (b) $0.2$, (c) $0.36$, (d) $0.58$, (e) $0.82$, and (f) $1$. CMD and PIMD results are presented for comparison.}
\label{fig:pcf}
\end{figure*}

Since the motion of \gls{li} is influenced by nearby oxygen atoms in the \gls{lg} electrolyte, we analyzed the pair correlation function (PCF) for Li–O pairs at different \gls{li}/\gls{g3} ratios, as shown in Fig.~\ref{fig:pcf}.  Regardless of \gls{li} concentration, a sharp peak at \( r_1\approx2.2 \)~\AA~ indicates a strong Li–O interaction. The oxygen atoms originate from both the solute \gls{litfsi} and the solvent \gls{g3}. \gls{li} interacts not only with oxygen atoms bonded to sulfur in solute \gls{litfsi} but also with those from the ether and glycol groups of solvent \gls{g3}.  

Additionally, a weak PCF peak appears at \( r_2\approx4.5 \)~\AA~, with its intensity increasing significantly as the \gls{li}/\gls{g3} ratio rises. This observation is in good agreement with CMD simulation results reported by Wang et al.~\cite{WANG2023100061}. The peak corresponds to the second coordination shell of \gls{li} and primarily originates from oxygen atoms in solute \gls{litfsi} molecules. In \gls{litfsi}, each sulfur atom forms bonds with two oxygen atoms, resulting in an O–S–O triplet. \gls{li} ions coordinate asymmetrically, favoring one side of the \gls{litfsi} molecular chain. This leads to a configuration where \gls{li} binds to proximal oxygen atoms in the triplets at \( r_1 \approx 2.2 \)~\AA~ and coordinates with distal oxygen atoms at \( r_2 \approx 4.5 \)~\AA~. These two peaks suggest local chemical ordering in the first and second Li–O coordination shells. This structural feature significantly influences the dynamic diffusion behavior of \gls{li}~\cite{doi:10.1021/acs.jpcb.6b09203}. Furthermore, the alignment of PCF distributions between CMD and PIMD (Fig.~\ref{fig:pcf}) indicates that quantum effects have a negligible impact on Li–O coordination.

\subsection{Viscosity}
\label{sec:viscosity}

Viscosity plays a crucial role in the diffusion behavior of \gls{li} in the \gls{lg} electrolyte. Therefore, we investigate the viscosity of electrolytes with \gls{lg} mole ratios ranging from $0$ to $1$. Viscosity can be essentially categorized into shear viscosity and longitudinal viscosity. Shear viscosity characterizes the resistance encountered by fluid layers during relative motion, while longitudinal viscosity quantifies the internal frictional resistance of the fluid during compression or expansion.  

The viscosity was calculated using the stress autocorrelation function,  
\begin{equation}
\eta_{\alpha \beta}(t)=\frac{1}{k_{\mathrm{B}} T V} \int_{0}^{t}\left\langle S_{\alpha \beta}(0) S_{\alpha \beta}\left(t^{\prime}\right)\right\rangle d t^{\prime},
\label{equation:viscosity}
\end{equation}
where $S_{\alpha \beta}$ represents the stress tensor components ($\alpha, \beta = x, y, z$), $t$ is the correlation time, $\langle \cdot \rangle$ denotes the ensemble average, $k_\text{B}$ is the Boltzmann constant, $T$ is the temperature, and $V$ is the system volume.  

The longitudinal viscosity is computed as the average of three independent diagonal components:  
\begin{equation}
\eta_{\text{L}} = \frac{\eta_{xx} + \eta_{yy} + \eta_{zz}}{3}.
\end{equation}
Similarly, the shear viscosity is obtained by averaging the corresponding off-diagonal components:  
\begin{equation}
\eta_\text{S} = \frac{\eta_{xy} + \eta_{yz} + \eta_{zx}}{3}.
\end{equation}
Finally, the bulk viscosity is expressed as a function of the longitudinal and shear viscosities:  
\begin{equation}
\eta_{\text{B}} = \eta_\text{L} - \frac{4}{3} \eta_\text{S}.
\end{equation}

\begin{figure}
\centering
\includegraphics[width=1\columnwidth]{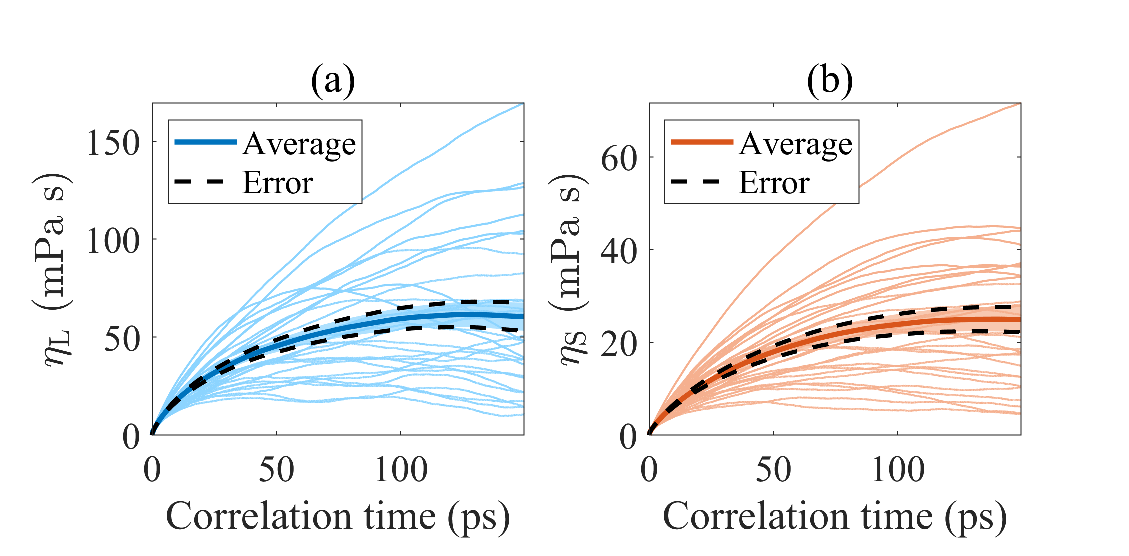}
\caption{CMD-derived (a) longitudinal and (b) shear viscosities versus correlation time for the \gls{lg} electrolyte at \gls{lg}~$=1$. Statistical averages were obtained from $30$ independent runs, with standard errors shown in each panel.}
\label{fig:viscosity-time}
\end{figure}

\begin{figure}
\centering
\includegraphics[width=0.8\columnwidth]{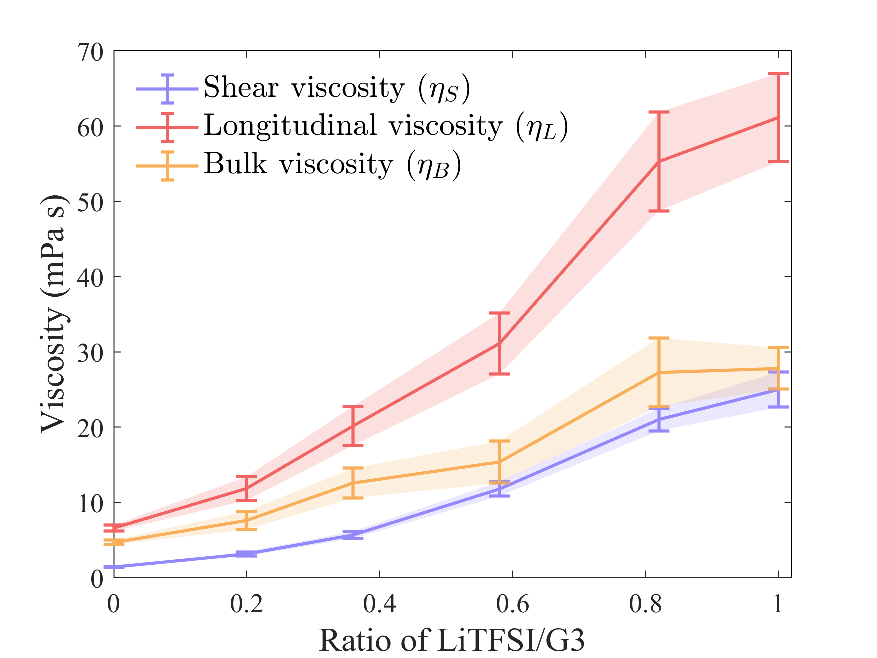}
\caption{Converged viscosity as a function of \gls{lg} concentration at $300$ K. Statistical averages were obtained from $30$ independent runs, with standard errors shown in each panel.}
\label{fig:viscosity-ratio}
\end{figure}

Figure~\ref{fig:viscosity-time} illustrates the time-correlated $\eta_\text{L}$ and $\eta_\text{S}$ curves for the electrolyte with \gls{lg}$=1$, where $30$ independent runs were performed to obtain statistical averages. The converged $\eta_\text{L}$ and $\eta_\text{S}$ values at different \gls{lg} ratios, along with the derived bulk viscosity $\eta_{\text{B}}$, are shown in Fig.~\ref{fig:viscosity-ratio}. A positive correlation is observed between viscosity and \gls{li} concentration~\cite{TOBISHIMA2004979}, indicating that \gls{li} enhances shear, longitudinal, and bulk viscosities. This effect is primarily attributed to the role of \gls{li} in the electrolyte, where solute and solvent oxygen atoms in the vicinity of \gls{li} interact strongly with it. Such interactions between the solute \gls{litfsi} and solvent \gls{g3} hinder the dynamic flow of the fluid.

\subsection{Self diffusion for lithium ions}

\begin{figure*}
\centering
\includegraphics[width=1.8\columnwidth]{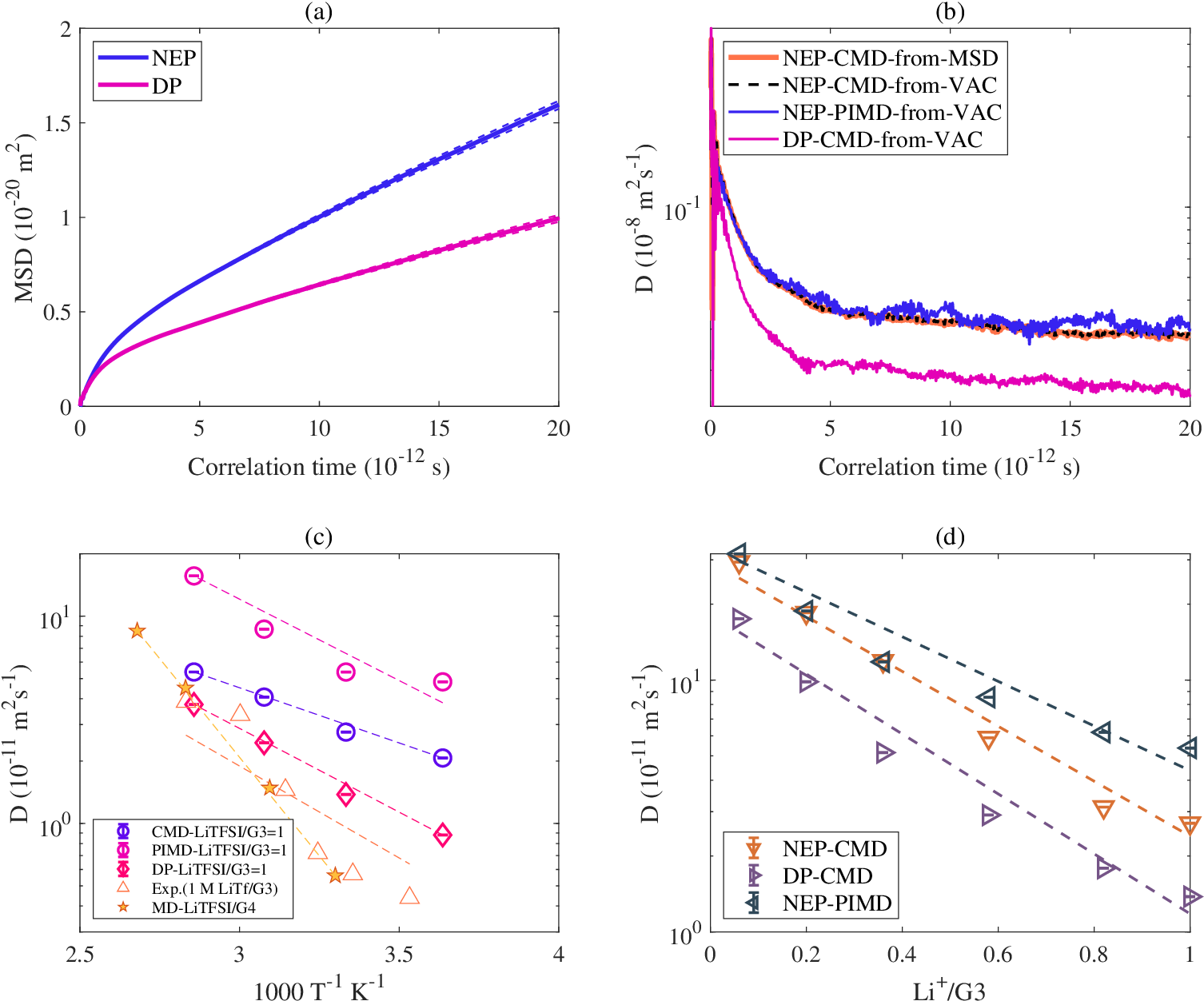}
\caption{Self-diffusion of \gls{li} in \gls{lg}. (a) CMD-calculated mean squared displacement versus correlation time for \gls{lg}$=0.0625$ at $300$ K. (b) Self-diffusion coefficients for \gls{lg}$=0.0625$ at $300$ K, derived from mean squared displacement and velocity autocorrelation methods. CMD-NEP, PIMD-NEP, and CMD-DP results are compared. (c) Self-diffusion coefficient as a function of inverse temperature for \gls{lg}$=1$. (d) Self-diffusion coefficient as a function of \gls{li} concentration at $300$ K. Experimental data for 1 M LiTf/G$_3$~\cite{C9TA03261D} and CMD simulations for LiTFSI/G$_4=1$~\cite{doi:10.1021/acs.jpcb.8b06913} are provided for reference.}
\label{fig:sdc}
\end{figure*}

In this section, we employ two distinct approaches to compute the self-diffusion coefficient ($D$). The first method derives $D$ from the time-dependent mean squared displacement (MSD) in the limit of large correlation time:

\begin{equation}
    \begin{aligned}
        D_{\mathrm{MSD}}(t)  &= \frac{1}{6} \lim_{t \rightarrow \infty} \frac{\mathrm{d}}{\mathrm{d}t} \mathrm{MSD}(t) \\
        &= \frac{1}{6} \lim_{t \rightarrow \infty} \frac{\mathrm{d}}{\mathrm{d}t} \langle [\mathbf{r}(t)-\mathbf{r}(0)]^2\rangle,
    \end{aligned}
\end{equation}
where $\mathbf{r}(t)$ denotes the position of \gls{li} at time $t$.

An alternative approach estimates $D$ using the velocity autocorrelation function (VAC):

\begin{equation}
    D_{\mathrm{VAC}}(t)=\frac{1}{3}\int_{0}^{t}\left\langle \mathbf{v}(0) \cdot \mathbf{v}(t')\right\rangle dt',
    \label{equation:D_VAC}
\end{equation}
where $\mathbf{v}$ represents the velocity of \gls{li}, and $\langle \cdot \rangle$ denotes the ensemble average.

Figure~\ref{fig:sdc} presents the self-diffusion behavior of \gls{li} in the electrolyte. We first computed the mean squared displacement (MSD) as a function of correlation time for \gls{lg}$=0.0625$, as shown in Fig.~\ref{fig:sdc}(a). Our NEP model predicts a higher MSD trend compared to the DP model in CMD simulations. This difference is attributed to the mass density of the electrolyte, as the DP model overestimates the density (see Fig.~\ref{fig:density}(b)), resulting in a denser \gls{lg} system where the diffusion of \gls{li} is hindered.  

The derivatives of $\mathrm{MSD}(t)$, corresponding to the self-diffusion coefficient $D_\mathrm{MSD}$, are plotted in Fig.~\ref{fig:sdc}(b) alongside the $D_\mathrm{VAC}$ values obtained from the velocity autocorrelation function in Eq.~\ref{equation:D_VAC}. Due to the steeper slope in Fig.~\ref{fig:sdc}(a), the MSD-derived $D_{\mathrm{MSD}}$ from our NEP model is higher than that predicted by the DP model. Moreover, the near-perfect alignment between the $D_\mathrm{MSD}(t)$ and $D_\mathrm{VAC}(t)$ curves confirms the equivalence of the two methods for computing self-diffusion in CMD simulations. Furthermore, the consistency between CMD and PIMD calculations in Fig.~\ref{fig:sdc}(b) indicates that quantum effects have a negligible impact on the self-diffusion behavior of \gls{li} in electrolytes with low \gls{li} concentrations.

By extracting the converged values of $D(t)$ at long correlation times, we further investigate the effects of temperature and \gls{li} concentration on the self-diffusion coefficient of \gls{li} in the \gls{lg} electrolyte. As shown in Fig.~\ref{fig:sdc}(c), \gls{li} exhibits a linear inverse correlation between $D$ and temperature, following a $D \propto T^{-1}$ dependence for \gls{lg}$=1$. As expected, higher temperatures facilitate diffusion, resulting in larger $D$ values.  The NEP-predicted $D \propto T^{-1}$ trend in the \gls{lg} system is in good agreement with DP-calculated results, previous MD simulations of LiTFSI/G$_4$ electrolytes~\cite{doi:10.1021/acs.jpcb.3c05612}, and experimental measurements of a $1$ M LiTf/G$_3$ system~\cite{C9TA03261D}.

In addition, \gls{li} exhibits a similar dependence on \gls{litfsi} concentration at $300$ K. As shown in Fig.~\ref{fig:sdc}(d), increasing the \gls{litfsi} ratio suppresses diffusion. Combined with the viscosity analysis in Sec.~\ref{sec:viscosity}, this trend suggests that a higher \gls{litfsi} mole fraction strengthens the interactions between \gls{li} and neighboring oxygen atoms in both solute \gls{litfsi} and solvent \gls{g3} (see Fig.~\ref{fig:pcf}), thereby hindering \gls{li} diffusion. Furthermore, self-diffusion exhibits visible quantum effects at higher \gls{li} concentrations, despite the negligible difference observed at lower concentrations. Moreover, the DP model consistently underestimates self-diffusion coefficients with respect to both temperature and \gls{li} concentration due to its overestimation of density.

\section{Conclusions}

In this study, we developed a high-accuracy neuroevolution potential (\gls{nep}) for the \gls{lg} (\gls{litfsi}--\gls{g3}) electrolyte and applied it to large-scale molecular dynamics simulations. By employing an active learning strategy, we constructed a dataset encompassing diverse electrolyte configurations, enabling the \gls{nep} model to achieve near-\gls{dft} accuracy while maintaining computational efficiency. The model was validated against benchmark \gls{dft} calculations, demonstrating excellent predictive performance for energies, forces, and virial stresses.

Using this \gls{nep} model, we systematically investigated the structural and transport properties of the electrolyte, including density, ion coordination, viscosity, and \gls{li} self-diffusion. The computed densities closely matched experimental measurements, with \gls{pimd} simulations further confirming the minor role of quantum effects in this system. Pair correlation function analysis revealed that \gls{li} coordination involves both solvent \gls{g3} and solute \gls{litfsi}, influencing its dynamic behavior. Viscosity calculations indicated that increasing \gls{litfsi} concentration strengthens ion-ion interactions, leading to higher viscosity and reduced \gls{li} mobility. Self-diffusion coefficients exhibited a clear dependence on both temperature and concentration, aligning well with experimental trends. These findings establish NEP as a robust and efficient framework for accurately modeling electrolyte behavior and gaining deeper insights into ion transport mechanisms.

\vspace{0.5cm}

\noindent{\textbf{Data availability:}}
    Complete input and output files for the \gls{nep} training and testing are freely available at a Zenodo repository \cite{cao_2025_data}. 

\begin{acknowledgments}    
    Y.S appreciates the support received from the Fundamental Research Funds for the Central Universities (FRF-TP-22-106A1). C.C. and P.Q. acknowledge the support from the National Key Research and Development Program of China (2021YFB3802104).
    We acknowledge the computational resources provided by High Performance Computing Platform of Beijing Advanced Innovation Center for Materials Genome Engineering.
\end{acknowledgments}

\bibliography{refs}

\end{document}